\documentclass[12pt, a4paper]{article}
\pdfoutput=1

\usepackage{cite}
\usepackage{amsmath,amssymb}
\usepackage{comment}
\usepackage{multirow}
\usepackage[utf8]{inputenc}
\usepackage{bm}
\usepackage{accents}
\usepackage{cite}

\usepackage{ifpdf}
\ifpdf        
  \usepackage{graphicx, hyperref, xcolor}     
\else     
  \usepackage[dvipdfmx]{graphicx, hyperref, xcolor}     
 \fi
\usepackage{subcaption}
 
\definecolor{rossoferrari}{HTML}{D9073D}
\definecolor{mediumblue}{HTML}{0000CD}
\hypersetup{
setpagesize=false,
bookmarksnumbered=true,%
bookmarksopen=true,%
colorlinks=true,%
linkcolor=rossoferrari,
urlcolor=mediumblue,
citecolor=mediumblue,
}

\usepackage[height=21.5cm,width=16.5cm,centering]{geometry}

\newlength{\dhatheight} 
\newcommand{\doublehat}[1]{%
    \settoheight{\dhatheight}{\ensuremath{\hat{#1}}}%
    \addtolength{\dhatheight}{-0.25ex}%
    \hat{\vphantom{\rule{1pt}{\dhatheight}}%
    \smash{\hat{#1}}}
}

\newcommand{\hyphen}{\,\mathchar`-\mathchar`-\,}

\begin{document}

\begin{titlepage}

\begin{center}

\hfill UT-18-24\\
\hfill KEK-TH-2083\\
\hfill DESY 18-180

\vskip .75in

{\Large \bf 
 Probing Electroweakly Interacting Massive Particles\\ \vspace{2mm}
 with Drell-Yan Process at 100 TeV Hadron Colliders
}

\vskip .75in

{\large
So Chigusa$^{(a)}$, Yohei Ema$^{(b, c)}$, and Takeo Moroi$^{(a)}$
}

\vskip 0.25in

$^{(a)}${\em Department of Physics, Faculty of Science,\\
The University of Tokyo,  Bunkyo-ku, Tokyo 113-0033, Japan}\\[.3em]
$^{(b)}${\em DESY, Notkestra{\ss}e 85, D-22607 Hamburg, Germany}\\[.3em]
$^{(c)}${\em KEK Theory Center, Tsukuba 305-0801, Japan}

\end{center}
\vskip .5in

\begin{abstract}

  There are many models beyond the standard model which include
  electroweakly interacting massive particles (EWIMPs), often in the
  context of the dark matter.  In this paper, we study the indirect
  search of EWIMPs using a precise measurement of the Drell-Yan cross
  sections at future $100\,{\rm TeV}$ hadron colliders.  It is
  revealed that this search strategy is suitable in particular for
  Higgsino and that the Higgsino mass up to about $1.3\,{\rm TeV}$ will be
  covered at $95\,\%$ C.L. irrespective of the chargino and neutralino
  mass difference.  We also show that the study of the Drell-Yan
  process provides important and independent information about every
  kind of EWIMP in addition to Higgsino.

\end{abstract}

\end{titlepage}


\renewcommand{\thepage}{\arabic{page}}
\setcounter{page}{1}
\renewcommand{\thefootnote}{\#\arabic{footnote}}
\setcounter{footnote}{0}

\section{Introduction}

There are many models that extend the standard model (SM) by
introducing electroweakly interacting massive particles (EWIMPs).  One
motivation to introduce them is the existence of the dark matter (DM)
in our universe.  For example, an electroweakino in the minimal
supersymmetric (SUSY) extension of the SM can be the lightest
supersymmetric particles (LSP) and are natural DM candidate.  In
particular, there are well motivated scenarios where the LSP is
Higgsino or wino, which transform as doublet and triplet under the
weak $SU(2)_L$ gauge symmetry, respectively; light Higgsino is
preferred to reduce the amount of the fine-tuning of the electroweak
scale as in the ``natural SUSY'' set up~\cite{Kitano:2005wc,
Brust:2011tb, Papucci:2011wy, Baer:2012uy}, while the so-called ``mini
split'' spectrum \cite{Wells:2003tf, Wells:2004di, ArkaniHamed:2004fb,
Giudice:2004tc, ArkaniHamed:2004yi, ArkaniHamed:2005yv} with anomaly
mediation \cite{Randall:1998uk, Giudice:1998xp} makes wino LSP.
Another example is the ``minimal dark matter'' (MDM)
scenario~\cite{Cirelli:2005uq, Cirelli:2007xd, Cirelli:2009uv}, in
which an additional EWIMP is introduced, whose stability is ensured
due to the large $SU(2)_L$ charge assignment.  In particular, 5-plet
fermion with hypercharge zero is a good DM candidate that escape from
the DM search experiments so far.

In order to search for EWIMPs, several different approaches are
adopted.  One approach is the direct production at the large hadron
collider (LHC).  The main strategy is to use the disappearing charged
track, which indicates a long-lived charged particle, the charged
components of the EWIMP in the current case.  Both ATLAS and CMS
collaborations announced a result of this method with the data of
$\sqrt{s} = 13\,{\rm TeV}$ LHC \cite{Aaboud:2017mpt,
  ATL-PHYS-PUB-2017-019, Sirunyan:2018ldc}.  The current lower bound
on the mass of the pure Higgsino-like (wino-like) state is
$152~(460)\,{\rm GeV}$ at $95\,\%$ C.L.  We can obtain a similar bound
for the MDM using the same method~\cite{Ostdiek:2015aga}.  In this
method, however, the bound strongly depends on the lifetime of the
charged component, which is sensitive to the mass difference between
the charged and the neutral components.  In particular, it is often
the case in the SUSY model that the Higgsino-like LSP and its charged
counterpart possess a non-negligible fraction of wino, which
significantly enhances the mass difference compared to the pure
Higgsino case.  In such a case, the lifetime of the charged component
is extremely short, making the disappearing track method challenging.
There is another option called mono-X search to search for a new
physics signal in general.  However, the corresponding bound is
usually very weak due to the large SM background and no bound is
imposed on Higgsino at $\sqrt{s} = 14\,{\rm TeV}$
LHC~\cite{Baer:2014cua}.

Another way of probing EWIMPs is DM search experiments, 
assuming that the EWIMPs are the dominant component of the DM.
Firstly, there exist several direct detection
experiments that utilize a scattering between the DM and the
nucleon~\cite{Akerib:2016vxi, Cui:2017nnn, Aprile:2018dbl}.  Wino is
the most promising target of these experiments, whose spin independent
scattering cross section with nucleon is almost mass independently
given as $\sigma_{\rm SI} \simeq 2.3 \times 10^{-47}\,{\rm
cm}^2$~\cite{Hisano:2010fy, Hisano:2012wm, Hisano:2015rsa,
Hill:2011be, Hill:2013hoa}, which is still an order of magnitude below
the current experimental limit.  It is unlikely to detect Higgsino
DM in this way since its small $SU(2)_L$ charge makes
scattering cross section comparable to or below the neutrino
floor~\cite{Hisano:2012wm}.  The detection of MDM may
also be difficult~\cite{Hisano:2011cs} since its possibly larger mass
of $\mathcal{O} (10)\,{\rm TeV}$ weakens the sensitivity of direct
detection experiments.

Secondly, a lot of effort is devoted to detecting cosmic rays resulting
from DM annihilation, namely the DM indirect
detection~\cite{Fermi-LAT:2016uux, Ahnen:2016qkx, Abdallah:2016ygi,
Gomez-Vargas:2013bea}.  Although the results suffer from some
astrophysical uncertainties, they have already excluded, for example,
wino with mass less than $400\,{\rm GeV}$ and also around $2\,{\rm
TeV}$~\cite{Bhattacherjee:2014dya}.
On the other hand, the corresponding Higgsino bound is weaker
and it has been probed only up to $350\,{\rm GeV}$~\cite{Krall:2017xij}
again due to the smallness of its $SU(2)_L$ charge.  For the MDM, 5-plet
fermion is analyzed as an example in \cite{Abdalla:2018mve} and the mass
less than $2\,~{\rm TeV}$ and several narrow regions are excluded.  Note
again that the EWIMPs must be the dominant component of the DM for these
approaches to be sensitive.

In order to acquire somewhat stronger bounds on the EWIMPs, independent
of the decay product or the lifetime and whether they are the dominant
component of the DM or not, it has been discussed that indirect search
of EWIMPs at collider experiments is useful~\cite{Alves:2014cda,
Gross:2016ioi, Farina:2016rws, Harigaya:2015yaa, Matsumoto:2017vfu}.  It
utilizes the EWIMP loop effect on various observables.  We pursue this
possibility and study the effects of EWIMPs on the oblique correction to
the electroweak gauge bosons.  In particular, in this paper, we study
the prospect of the indirect search method at future $100\,{\rm TeV}$
hadron colliders such as FCC-hh at CERN~\cite{Mangano:2016jyj,
Contino:2016spe, Golling:2016gvc} or SppC in
China~\cite{CEPC-SPPCStudyGroup:2015csa, CEPC-SPPCStudyGroup:2015esa}.
We concentrate on the Drell-Yan process that has two leptons in the
final state since it provides a very clean signal without any hadronic
jets at least from the final state particles.  We will show that it
provides a comparable or better experimental reach for Higgsino compared
to the direct production search of EWIMPs at future
colliders~\cite{Low:2014cba, Cirelli:2014dsa, Han:2018wus,
Mahbubani:2017gjh}.  This method also provides independent and
additional information about wino and the MDM.

The rest of the paper is organized as follows.  In
Sec.~\ref{sec:ewimp}, we summarize the systematic way of calculating
EWIMP effect on the Drell-Yan process.  In Sec.~\ref{sec:analysis}, we
describe our statistical procedure to determine the bounds 
and show the corresponding results. 
We also briefly comment on the mass determination of EWIMPs.
In Sec.~\ref{sec:conclusion}, we summarize our results.

\section{EWIMP effect on the Drell-Yan process}
\label{sec:ewimp}

We assume that all the beyond SM particles except for EWIMPs are much
heavier than $\mathcal{O}(1)\,{\rm TeV}$.  Then, new physics affects
the Drell-Yan process only through the vacuum polarization of the
electroweak gauge bosons with EWIMPs in the loop.  After integrating
out EWIMPs,\footnote{Contrary to the usual effective field theory
  approach, our prescription can also be applied to the energy scale
  $Q \gtrsim m$ since we do not perform a series expansion of the loop
  function $f$ in Eq.~\eqref{eq_lag}.} this effect can be expressed as
\begin{align}
 \mathcal{L}_{\rm eff} = \mathcal{L}_{\rm SM} + C_2 W_{\mu \nu}^a
 f\left(-\frac{D^2}{m^2}\right) W^{a\mu\nu} + C_1 B_{\mu\nu}
 f\left(-\frac{\partial^2}{m^2}\right) B^{\mu\nu},\label{eq_lag}
\end{align}
where $\mathcal{L}_{SM}$ is the SM Lagrangian, $D$ is a covariant
derivative acting on $W^{a\mu\nu}$, $m$ is the EWIMP mass,\footnote{Here
we neglect a small mass splitting among $SU(2)_L$ $n$-plet.}
and $W_{\mu\nu}^a$ and $B_{\mu\nu}$ are field strength associated with
the $SU(2)_L$ and $U(1)_Y$ gauge group, respectively.
The function $f(x)$ is a
loop function defined as
\begin{align}
 f(x) = \begin{cases}
	 \displaystyle{\frac{1}{16\pi^2} \int_0^1 dy\, y(1-y) \ln (1 -
	 y(1-y)x - i0)} & {\rm (Fermion)},\\[5mm]
	 \displaystyle{\frac{1}{16\pi^2} \int_0^1 dy\, (1-2y)^2 \ln (1-
	 y(1-y)x - i0)} & {\rm (Scalar)},
	\end{cases}
\end{align}
where the first (second) line corresponds to a fermionic (scalar) EWIMP, respectively.
The coefficients $C_1$ and $C_2$ for an $n$-plet EWIMP with hypercharge $Y$ 
are given by
\begin{align}
 C_1 &= \frac{\kappa}{8} n Y^2 {g'}^2,\\
 C_2 &= \frac{\kappa}{96} (n^3 - n) g^2,
\end{align}
where $g'$ and $g$ are the $U(1)_Y$ and $SU(2)_L$ gauge couplings,
and $\kappa = 1, 2, 8, 16$ 
for a real scalar, a complex scalar, a Weyl or a
Majorana fermion, and a Dirac fermion, respectively.  In particular,
$(C_1, C_2) = ({g'}^2, g^2)$ for Higgsino and $(C_1, C_2) = (0,
2g^2)$ for wino.

The parton level matrix element for the Drell-Yan process
$q(p)~\bar{q}(p') \to \ell^{-}(k)~\ell^{+}(k')$ at the leading order
(LO) within the SM is
\begin{align}
 \mathcal{M}_{\rm SM} = \sum_{V = \gamma, Z} \frac{\left[ \bar{v}(p')
 \gamma^\mu \Gamma_q^{(V)} u(p) \right] \left[ \bar{u}(k) \gamma_\mu
 \Gamma_\ell^{(V)} v(k') \right]}{m_{\ell\ell}^2 - m_V^2},\label{eq_m_sm}
\end{align}
where $m_{\ell\ell}$ and $m_V$ are the final state lepton invariant
mass and the electroweak gauge boson mass, respectively.  Here,
$\{\Gamma_f^{(\gamma)}, \Gamma_f^{(Z)}\} = \{e Q_f, g_Z (v_f - a_f
\gamma_5)\}$ with $e = g s_W$ and $g_Z = g / c_W$, where $s_W \equiv
\sin \theta_W$, $c_W \equiv \cos \theta_W$ with $\theta_W$ being the
weak mixing angle.  The parameters ${Q_f, v_f, a_f}$ for the up-type
quarks, the down-type quarks, and the leptons are given by $\{2/3,
1/4-2s_W^2/3, 1/4\}$, $\{-1/3, -1/4+s_W^2/3, -1/4\}$, and $\{-1,
-1/4+s_W^2, -1/4\}$, respectively.  From Eq.~\eqref{eq_lag}, we can
also calculate the leading contribution to the matrix element from
EWIMPs as
\begin{align}
 \mathcal{M}_{\rm EWIMP} = \sum_{V, V' = \gamma, Z} C_{V V'} m_{\ell\ell}^2
 f\left(\frac{m_{\ell\ell}^2}{m^2}\right) \frac{\left[ \bar{v}(p') \gamma^\mu \Gamma_q^{(V)}
 u(p) \right] \left[ \bar{u}(k) \gamma_\mu \Gamma_\ell^{(V')} v(k')
 \right]}{(m_{\ell\ell}^2-m_V^2) (m_{\ell\ell}^2-m_{V'}^2)},\label{eq_m_ewimp}
\end{align}
where $C_{\gamma \gamma} = (C_1 c_W^2 + C_2
s_W^2)/2$, $C_{\gamma Z} = C_{Z \gamma} = (C_2 - C_1) s_W c_W/2$, and $C_{Z
Z} = (C_1 s_W^2 + C_2 c_W^2)/2$.

In order to calculate the differential cross section $d\sigma / dm_{\ell\ell}$, we
need to integrate over the initial parton energy distributions and the
final state phase space.  Let $dL_{ab} / dm_{\ell\ell}$ be a luminosity
function for a fixed $m_{\ell\ell}$,
\begin{align}
 \frac{d L_{ab}}{dm_{\ell\ell}} \equiv \frac{1}{s} \int_0^1 dx_1
 dx_2~f_a(x_1) f_b(x_2) \delta\left(\frac{m_{\ell\ell}^2}{s} - x_1 x_2\right),
\end{align}
where $a,b$ denote species of initial partons, $\sqrt{s}$ is a center of mass
energy of the proton collision ($\sqrt{s} = 100\,{\rm TeV}$ in our case),
and $f_a(x)$ is a parton distribution function (PDF) of the given parton
$a$.  We also define $d \Pi_{\rm LIPS}$ as a Lorentz invariant phase
space for the two body final state.  Then, with
Eqs.~\eqref{eq_m_sm} and~\eqref{eq_m_ewimp},
we obtain the differential cross section as
\begin{align}
 \frac{d \sigma_{\rm SM}^{ab}}{dm_{\ell\ell}} &= \int d\Pi_{\rm LIPS}~\left|
 \mathcal{M}_{\rm SM} \left( q_a \bar{q}_b \to \ell^{-} \ell^{+} \right)
 \right|^2,\label{eq_sij_sm}\\
 \frac{d \sigma_{\rm EWIMP}^{ab}}{dm_{\ell\ell}} &= \int d\Pi_{\rm LIPS}~2 \Re
 \left[ \mathcal{M}_{\rm SM} \mathcal{M}_{\rm EWIMP}^{*} \left( q_a
 \bar{q}_b \to \ell^{-} \ell^{+} \right) \right],\label{eq_sij_ewimp}\\
 \frac{d \sigma_{\rm SM}}{dm_{\ell\ell}} &= \sum_{a,b} \frac{dL_{ab}}{dm_{\ell\ell}} \frac{d
 \sigma_{\rm SM}^{ab}}{dm_{\ell\ell}},\label{eq_sig_sm}\\
 \frac{d \sigma_{\rm EWIMP}}{dm_{\ell\ell}} &= \sum_{a,b} \frac{dL_{ab}}{dm_{\ell\ell}}
 \frac{d \sigma_{\rm EWIMP}^{ab}}{dm_{\ell\ell}}.\label{eq_sig_ewimp}
\end{align}
Here, $d\sigma_{\rm SM} / dm_{\ell\ell}$ is the SM cross section of
the Drell-Yan process, while $d\sigma_{\rm EWIMP} / dm_{\ell\ell}$ is
an EWIMP contribution to the cross section.  Thus, with introducing the
parameter $\mu$, we denote the total cross section as
\begin{align}
 \frac{d\tilde{\sigma}}{dm_{\ell\ell}} = 
 \frac{d\sigma_{\rm SM}}{dm_{\ell\ell}} 
 + \mu \frac{d\sigma_{\rm EWIMP}}{dm_{\ell\ell}}.
 \label{eq_diffcrosssection}
\end{align}
Obviously, $\mu=0$ corresponds to the SM, while $\mu=1$ corresponds to
the SM$+$EWIMP model.  Hereafter, we use
\begin{align}
 \delta_\sigma^{ab} (m_{\ell\ell}) \equiv \frac{d\sigma_{\rm EWIMP}^{ab}
 / dm_{\ell\ell}}{d\sigma_{\rm SM}^{ab} /
 dm_{\ell\ell}},\label{eq_dsigma}
\end{align}
which parameterizes the amount of the correction.  Note that this ratio
remains unchanged even if we take into account the next-to-leading order
(NLO) QCD effect because the EWIMPs affect the cross sections only
through the vacuum polarization.\footnote{When the NLO QCD effect is
included, one of the initial partons can be gluon with the real emission
of one jet in the final state.  However, we can easily see that
$\delta_\sigma^{ug}$ = $\delta_\sigma^{uu}$ and so on.}

\begin{figure}[t]
  \centering
  \begin{subfigure}{0.495\linewidth}
   \includegraphics[width=\linewidth]{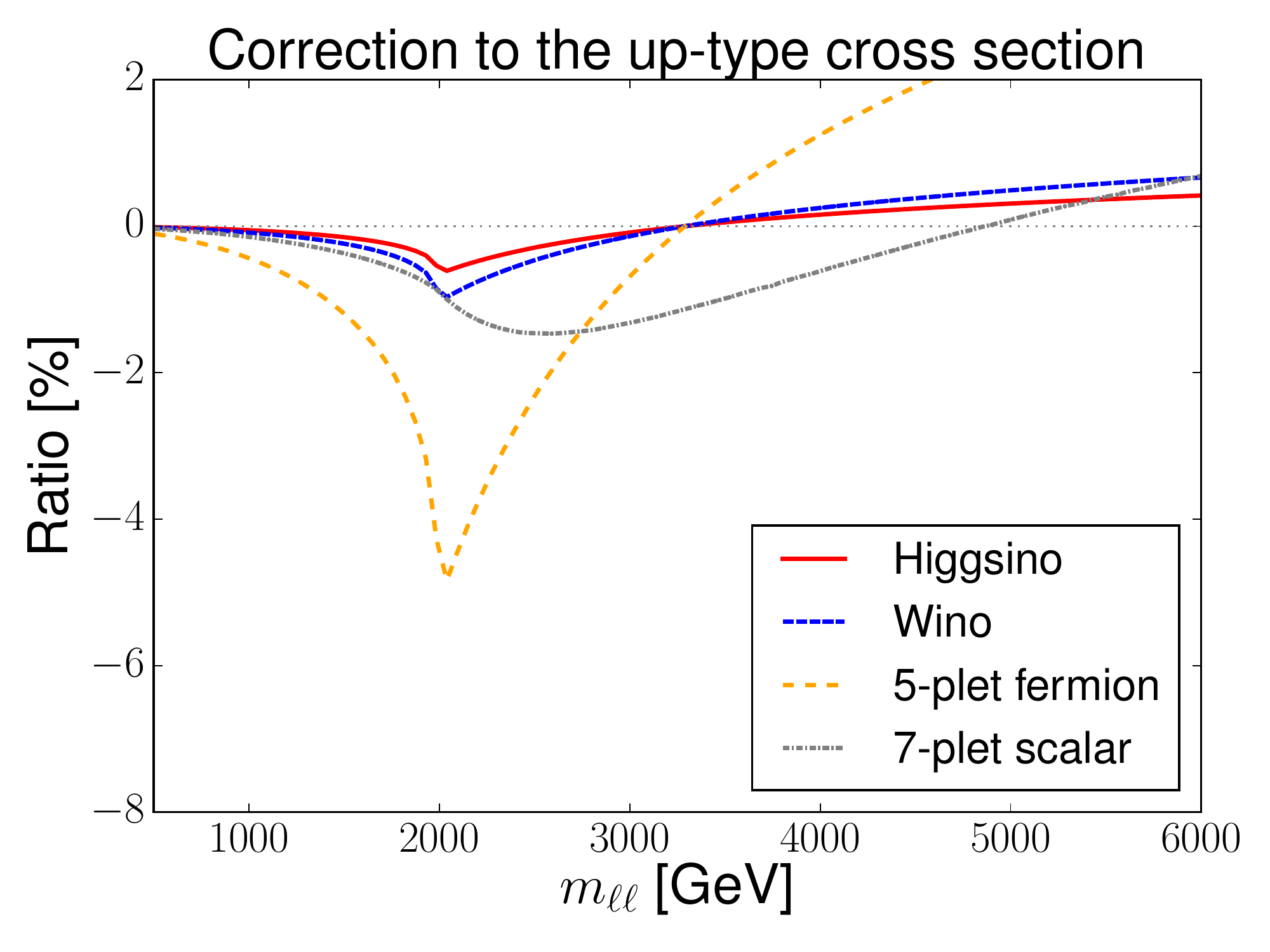}
  \end{subfigure}
  \begin{subfigure}{0.495\linewidth}
   \includegraphics[width=\linewidth]{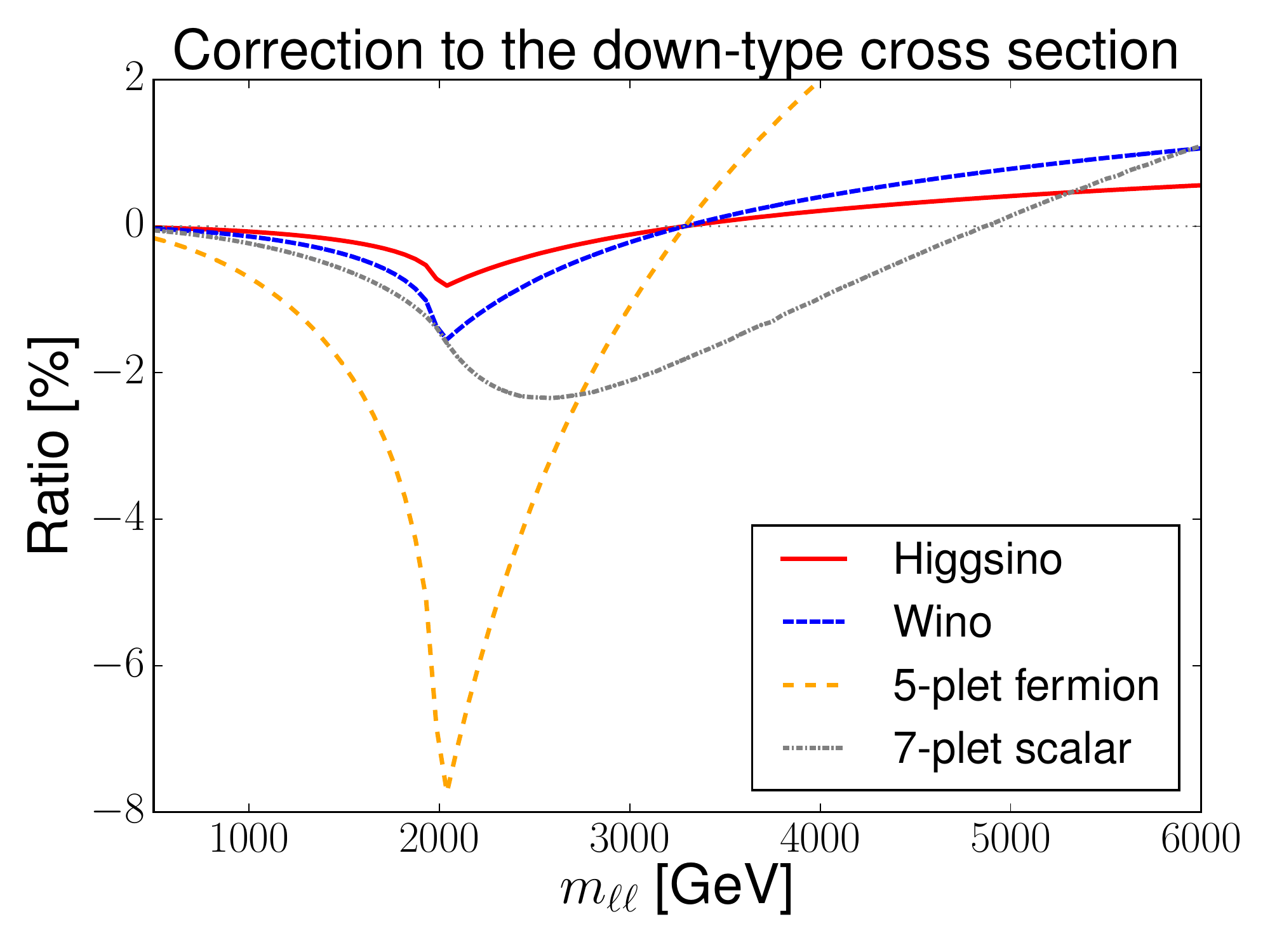}
  \end{subfigure}
 \caption{Correction to the differential cross section normalized by
 the SM contribution as a function of the lepton invariant mass
 $m_{\ell\ell}$.  The $y$-axis corresponds to $\delta_\sigma^{uu}$
 (left) and $\delta_\sigma^{dd}$ (right).  Four examples of EWIMPs are
 shown: Higgsino, wino, 5-plet fermion, and 7-plet scalar.  All the
 masses are fixed to be $1\,{\rm TeV}$.
 } \label{fig_sig_ratio}
\end{figure}

In Fig.~\ref{fig_sig_ratio}, we show $\delta_\sigma^{uu}$ (left) and
$\delta_\sigma^{dd}$ (right) as a function of $m_{\ell\ell}$.  The red,
blue, orange, and gray lines correspond to Higgsino, wino, 5-plet
fermion, and 7-plet scalar\footnote{In~\cite{DiLuzio:2015oha,
DelNobile:2015bqo}, it is pointed out that 7-plet scalar decays too fast
due to the higher dimensional operators to be a dark matter candidate.
Below, we analyze it and show corresponding results just as an example
of the scalar EWIMP.}, respectively, with their masses fixed to be
$1\,{\rm TeV}$.  Note that the lines sharply bend at $m_{\ell\ell} =
2m$, which is clearer for the fermionic case.  These bends are due to
the fact that the pair production channel of the EWIMPs opens just above
$m_{\ell\ell} = 2m$.  It is this characteristic shape that plays an
important role for the discrimination of the EWIMP signal from various
systematic errors (see Sec.~\ref{sec_statistical}) and also for the
determination of the EWIMP mass (see Sec.~\ref{sec_mass}).

\section{Analysis}
\label{sec:analysis}

\subsection{Event generation}
\label{sec_event}

In our analysis, we first generate events for the Drell-Yan process
within the SM.  We assume the center of mass energy $\sqrt{s} =
100\,{\rm TeV}$ and the integrated luminosity $\mathcal{L} = 30\,{\rm
  ab}^{-1}$.  For the event generation, we use
\verb|MadGraph5_aMC@NLO|~\cite{Alwall:2011uj, Alwall:2014hca} for the
hard process, followed by parton shower and hadronization with
\verb|Pythia8|~\cite{Sjostrand:2014zea}, and detector simulation with
\verb|Delphes3|~\cite{deFavereau:2013fsa}.  We use \verb|NNPDF2.3QED|
with $\alpha_s(M_Z) = 0.118$~\cite{Ball:2013hta} as an input set of
PDFs.  As renormalization and factorization scales, we take the
geometric mean of lepton transverse momenta (which we call $Q$).  We
take into account the NLO QCD effect by the \verb|[QCD]| option of
\verb|MadGraph5_aMC@NLO| since it enhances the cross section roughly
by a factor of $2$ compared to the LO calculation.\footnote{
  This large enhancement implies that the next-to-next-to-leading order
  QCD effect may also have a non-negligible effect on the cross section,
  and its calculation remains as a future task.  However, due to its
  smooth dependence on $m_{\ell\ell}$, it may not much affect the
  detection reach of the EWIMPs.  See Sec.~\ref{sec_statistical} for the
  details.
} 
Events with $\ell = e,\mu$ are generated
within the range of $500\,{\rm GeV} < m_{\ell\ell} < 15000\,{\rm GeV}$
and divided into $145$ bins with $100\,{\rm GeV}$ width each.  
We use the detector card for the FCC-hh detector installed in
\verb|Delphes3| by default.  The main detector effects are 
i) reducing the total number of events by about $20\,\%$
because of the lepton miss-identification and ii) smearing lepton
momenta by a unit of $\mathcal{O}(1- 10)\,{\rm GeV}$ depending on
the original momentum.

The EWIMP effect can be incorporated by using $\delta_\sigma^{ab}$
defined in Eq.~\eqref{eq_dsigma} and rescaling the SM event.  The
event number in the SM $+$ EWIMP model in the bin $m_{\ell\ell}^{\rm
  min} < m_{\ell\ell} < m_{\ell\ell}^{\rm max}$ is theoretically
estimated as
\begin{align}
  N_{{\rm SM}+{\rm EWIMP}} = \sum_{m_{\ell\ell}^{\rm min} <
    m_{\ell\ell}^{\rm obs} < m_{\ell\ell}^{\rm max}} 
  \left[
    1 + \delta_\sigma^{ab} (m_{\ell\ell}^{\rm true}) 
  \right],
  \label{eq_n_tot}
\end{align}
where the sum runs over all events whose lepton invariant mass
$m_{\ell\ell}^{\rm obs}$ observed by the detector simulation falls
into the bin.\footnote{In the actual analysis, sufficient number of
  events are generated and each event is rescaled by some weight $w$
  in order to adjust the total event number to the corresponding value
  under $30\,{\rm ab}^{-1}$.  In this case, $w$ should be multiplied
  to the right-hand side of Eq.~\eqref{eq_n_tot}.  } Note that the
true invariant mass $m_{\ell\ell}^{\rm true}$ (and $a,b$), which is
extracted from the hard process information, is used for calculation
of $\delta_\sigma^{ab} (m_{\ell \ell}^{\rm true})$ since
$m_{\ell\ell}^{\rm true} \neq m_{\ell\ell}^{\rm obs}$ due to the
detector effect in general.

\subsection{Statistical treatment}
\label{sec_statistical}

In order to obtain the experimental detection reach, we use the
profile likelihood method.  With the cross section given in Eq.\
\eqref{eq_diffcrosssection}, we calculate the theoretically expected
number of events $\tilde{x}_i (\mu)$ in the $i$-th bin by substituting
$\delta_\sigma^{ab} \to \mu \delta_\sigma^{ab}$ in
Eq.~\eqref{eq_n_tot}.  

Up to now, we have evaluated the expected number of events without
taking into account systematic errors.  However, in the actual
experiment, the number of events may be affected, for example, by the
uncertainties of the integrated luminosity and the beam energy, choices
of the renormalization scale, the factorization scale, and PDF, detector
efficiency and acceptance, and so on.  In order to take account of the
systematic errors in association with these uncertainties, we introduce
nuisance parameters $\bm{\theta}=\{\theta_\alpha\}$, by which some (or
all) of the uncertainties might be absorbed.  We express our theoretical
prediction of the number of events in the $i$-th bin including the
systematic errors as
\begin{align}
  \tilde{x}_i(\bm{\theta}, \mu) &\equiv \tilde{x}_i(\mu)
  f_{\mathrm{sys},i}(\bm{\theta}),
\end{align}
where $f_{\mathrm{sys},i}(\bm{\theta})$ is some function of the nuisance
parameters that has a property $f_{\mathrm{sys},i} (\bm{0}) = 1$.  The
explicit form of $f_{\mathrm{sys},i}$ used in our analysis will be given
below.

In order to discuss the detection reach, we define the following test
statistic; once the number of events are experimentally measured, we
introduce \cite{Cowan:2010js}
\begin{align}
  q_0 = -2\ln \frac{L(\bm{x}; \doublehat{\bm{\theta}},
 \mu=0)}{L(\bm{x}; \hat{\bm{\theta}}, \hat{\mu})},
 \label{q(theta)}
\end{align}
where $\bm{x} = \{ x_i \}$ with $x_i$ being the observed number of
events in the $i$-th bin and $\doublehat{\bm{\theta}}$ and
$\{\hat{\bm{\theta}}, \hat{\mu}\}$ are the best fit values of
$\bm{\theta}$ (and $\mu$) which maximize $L(\bm{x}; \bm{\theta},
\mu=0)$ and $L(\bm{x}; \bm{\theta}, \mu)$, respectively.  The
likelihood function is defined as
\begin{align}
  L(\bm{x}; \bm{\theta}, \mu) &\equiv 
  L_{\bm{\theta}} (\bm{x}; \mu)
  L'(\bm{\theta}; \bm{\sigma}),
  \\
    L_{\bm{\theta}}(\bm{x}; \mu) &\equiv \prod_i
    \exp\left[-\frac{(x_i - \tilde{x}_i(\bm{\theta},\mu))^2}{2\tilde{x}_i(\bm{\theta},\mu)}
    \right],
  \\
  L'(\bm{\theta}; \bm{\sigma}) &\equiv 
  \prod_\alpha
  \exp \left[ -\frac{\theta_\alpha^2}{2\sigma_\alpha^2} \right].
  \label{eq_L_theta}
\end{align}
Here, $L'(\bm{\theta}; \bm{\sigma})$ is the likelihood function for
the nuisance parameters with $\bm{\sigma}$ (which we call
``variance'') parameterizing the possible sizes of $\bm{\theta}$ in
the SM.  
The limit $\bm{\sigma} \rightarrow \bm{0}$ corresponds to 
an optimistic case without any systematic errors.
We evaluate $q_0$ by taking $\bm{x} = \{\tilde{x}_i
(\mu=1)\}$ to estimate the detection reach of the EWIMPs in the
following.  (For such a choice, $\hat{\bm{\theta}} = \bm{0}$ and
$\hat{\mu} = 1$.) We identify $\sqrt{q_0} \simeq 1.96$ and $\sqrt{q_0}
= 5$ as the reach at $95\,\%$ C.L. and $5\sigma$ levels, respectively.
If we instead study the exclusion prospect, we may take $\bm{x} =
\{\tilde{x}_i (\mu=0)\}$.  Because the difference between the number
of events for $\mu=0$ and $\mu=1$ is numerically small in the present
case, the detection reach we will show in the following can also be
regarded as the exclusion reach of EWIMPs with good accuracy (see
Eqs.~(12) and (14) of Ref.~\cite{Cowan:2010js}).

Now we move on to the details of our treatment of the systematic
errors.  We do not know the best choice of
$f_{\mathrm{sys},i}(\bm{\theta})$ before the start of the experiment.
In this paper, we adopt what is known to work for the experimental
data~\cite{Aaltonen:2008dn} and take the functional form of
$f_{\mathrm{sys},i}(\bm{\theta})$ as
\begin{align}
  f_{\mathrm{sys},i}(\bm{\theta}) &= e^{\theta_1} (1 + \theta_2 p_i) p_i^{(\theta_3 +
    \theta_4 \ln p_i + \theta_5 \ln^2p_i)},\label{eq_ftheta}
\end{align}
where $p_i=2m_{\ell \ell, i} / \sqrt{s}$ with $m_{\ell \ell, i}$ being
the median of the $i$-th bin lepton invariant mass.  We will check
that the effects of the systematic errors of our concern can be
absorbed in $\bm{\theta}$ with this form of
$f_{\mathrm{sys},i}(\bm{\theta})$.

In order to estimate the variances of the nuisance parameters
$\bm{\sigma} = \{\sigma_\alpha\}$, we first perform the fit of the
results to potential sources of systematic errors using Eq.\
\eqref{eq_ftheta}.  We prepare two data sets (i.e., number of events
in each bin) $\bm{y} = \{y_i\}$ and $\bm{y'} = \{y_i'\}$, both of
which are obtained assuming the SM only hypothesis.  Here, $\bm{y}$ is
given by the original set up, while $\bm{y'}$ is calculated with
varying one of the followings:
\begin{enumerate}
\item Luminosity ($\pm5\,\%$),
\item Renormalization scale ($2Q$ and $Q/2$),
\item Factorization scale ($2Q$ and $Q/2$),
\item PDF choice.
\end{enumerate}
Then, we determine the best fit values of $\bm{\theta}$ by minimizing
the following $\chi^2$ variable:
\begin{align}
  \chi^2 &\equiv \sum_{i} \frac{(y'_i - \tilde{y}_i
    (\bm{\theta}))^2}{\tilde{y}_i(\bm{\theta})},\\  
  \tilde{y}_i(\bm{\theta}) &\equiv y_i f_{\mathrm{sys},i}(\bm{\theta}).
\end{align}
The best fit values are added in quadrature to estimate the variances.
The effect of the luminosity uncertainty is estimated by varying it by
$\pm5\,\%$.  The effects of renormalization and factorization scales
uncertainties are studied by varying them to $2Q$ and to $Q/2$.  The
effect of the PDF choice is studied by the ``systematics'' module that
is a built-in program of \verb|MadGraph5_aMC@NLO|.  We have checked
that, for the sources of the systematic errors listed above, $\bm{y'}$ is
well fitted by our choice of $f_{\mathrm{sys},i}(\bm{\theta})$.  In
Tab.~\ref{tab_sys_5}, we show the variances of $\bm{\theta}$ for each
source of the systematic error.  Assuming that all sources of the
systematic errors are uncorrelated, we combine all the systematic
errors; the result is shown in the last line denoted as ``Total'', which
is used as $\bm{\sigma}$ in Eq.~\eqref{eq_L_theta} when we include the
systematic errors.

\begin{table}[t]
  \centering
  \begin{tabular}{|c|c|c|c|c|c|}
  \hline
  Sources of systematic errors & $\sigma_1$ & $\sigma_2$ &
	       $\sigma_3$ & $\sigma_4$ & $\sigma_5$ \\ \hline
  \hline
  Luminosity: $\pm 5\,\%$ & $0.07$ & $0$ & $0$ & $0$ & $0$ \\ \hline
  Renormalization scale: $2Q, Q/2$ & $0.6$ & $0.9$ & $0.4$ & $0.08$ &
		       $0.006$ \\ \hline
  Factorization scale: $2Q, Q/2$ & $0.5$ & $0.7$ & $0.3$ & $0.07$ &
		       $0.007$ \\ \hline
  PDF choice & $0.4$ & $0.7$ & $0.3$ & $0.06$ & $0.004$ \\ \hline
  \hline
  Total & $0.9$ & $1.3$ & $0.5$ & $0.1$ & $0.01$ \\ \hline
  \end{tabular}
 \caption{Standard deviation of the parameters $\bm{\theta}$ derived
 through the fit of various systematic errors.  In the last line,
 the standard deviations are combined for each parameter assuming that all
 the errors are independent with each other.}  \label{tab_sys_5}
\end{table}

Several comments on other possible sources of systematic errors are in
order.  As for the beam energy error, we could not generate events at
NLO due to the lack of sufficient computational power.  Instead, we
checked at LO that the corresponding values of $\bm{\sigma}$ (assuming
that the uncertainty of the beam energy is $1\,\%$) are small enough,
and hence we simply ignored it.  Two of the remaining sources are the
pile-up effect and the underlying event, but they may be thought of as
negligible since we are focusing on the very clean signal of two
energetic leptons.  Another one is the effect of background processes
which is not considered in our analysis.  It is in principle possible
to estimate its effect through the simulation and improve the analysis
but here we just leave it as a future task.  Yet another source is the
error in the simulation of detector effect which can not be treated in
our procedure.  Related to this, we note here that a smooth change of
the event number in general, possibly including the uncertainty
related to the detector effect, could be absorbed by a minimization
procedure using some fit function like in Eq.~\eqref{eq_ftheta}.  On
the other hand, as we will discuss below, the EWIMP signal can not be
fully absorbed by the fit because of the sharp bend we mentioned
before.

\subsection{Detection reach}

\begin{figure}[t]
  \centering
  \begin{subfigure}{0.495\linewidth}
  \centering
   \includegraphics[width=\linewidth]{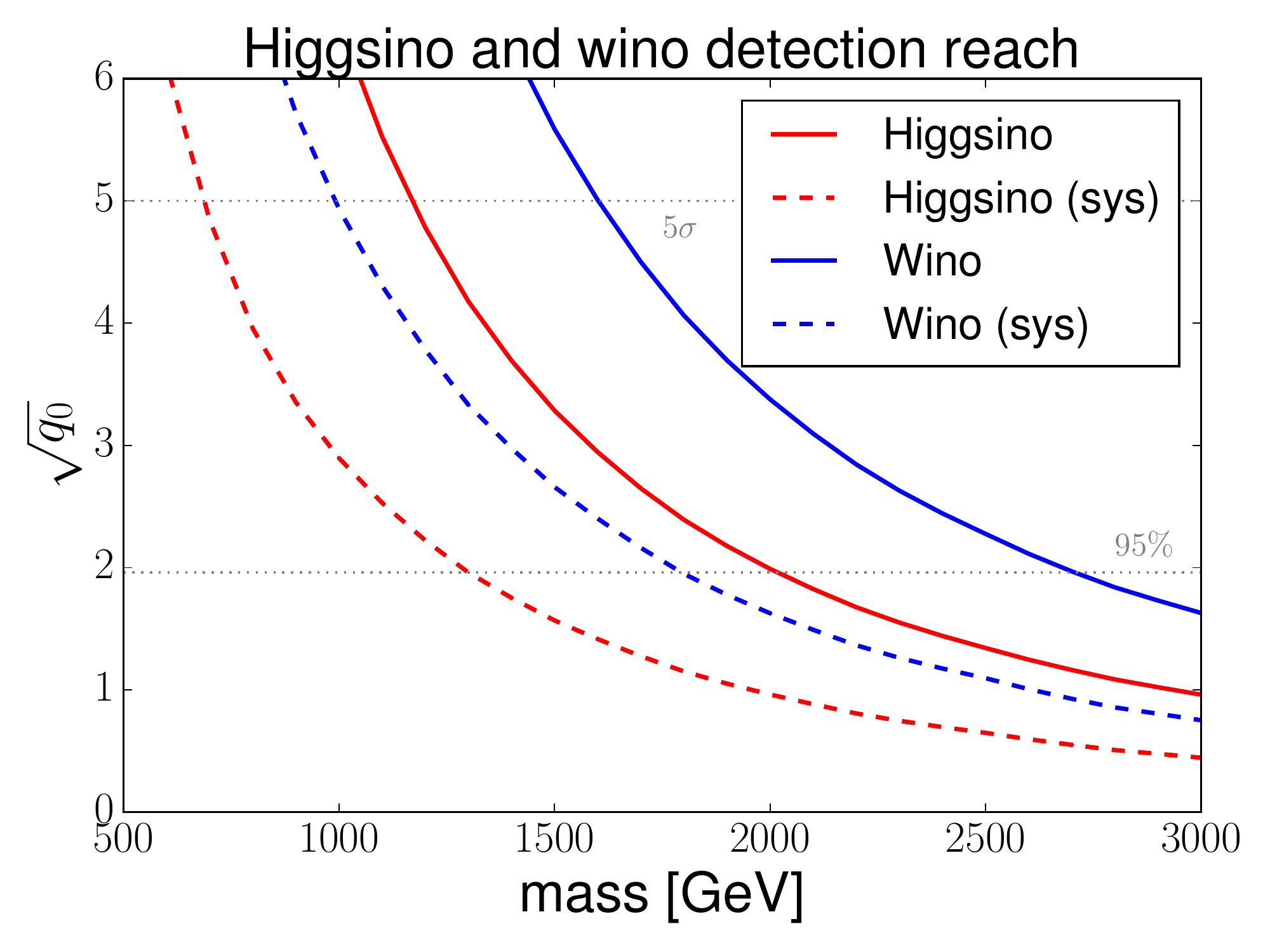}
  \caption{Higgsino and wino}
  \label{subfig:wh}
  \end{subfigure}
  \begin{subfigure}{0.495\linewidth}
  \centering
   \includegraphics[width=\linewidth]{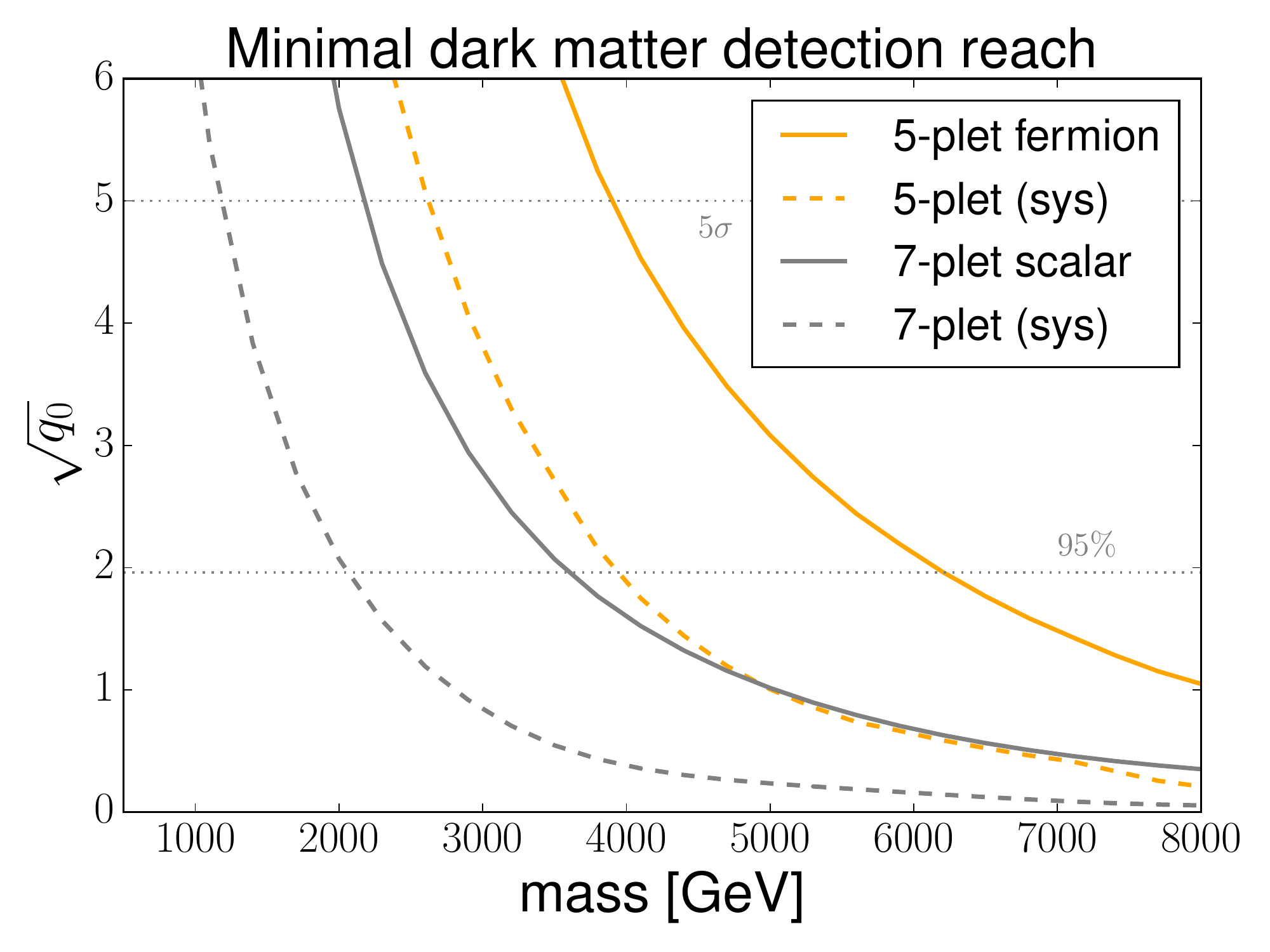}
  \caption{5-plet and 7-plet}
  \label{subfig:57}
  \end{subfigure}
 \caption{EWIMP detection reach with $30\ {\rm ab}^{-1}$ data of $100\
 {\rm TeV}$ hadron colliders.  The results with Higgsino and wino are shown
 in the left panel, while those with the MDMs are shown
 in the right panel.
 The solid lines are the optimistic one ($\bm{\sigma} \rightarrow \bm{0}$), 
 while the dashed lines are with the systematic errors.}
 \label{fig_chi2mass}
\end{figure}

Now we show the detection reach of EWIMPs at future $100\,{\rm TeV}$
hadron colliders.  In Fig.~\ref{fig_chi2mass}, we plot the value of
$\sqrt{q_0}$ as a function of the EWIMP mass. The solid and dashed
lines correspond to the optimistic detection reach without systematic
errors (i.e.  $\bm{\sigma} \to \bm{0}$) and the reach
obtained using the five parameters fit, respectively.  The red and
blue lines show the result with Higgsino and wino in
Fig.~\ref{subfig:wh}, while the orange and gray lines are the result
with 5-plet fermion and 7-plet scalar in Fig.~\ref{subfig:57}.

Let us start with the detection reach of Higgsino and wino in
Fig.~\ref{subfig:wh}.  We can see that the $5\sigma$ discovery reach is
about $1.2\,{\rm TeV}$ and $1.6\,{\rm TeV}$, while the $95\,\%$ lines
are $2.0\,{\rm TeV}$ and $2.7\,{\rm TeV}$ for Higgsino and wino in the
case without systematic errors.  Taking into account the systematic
errors with $\bm{\sigma}$ given in Tab.~\ref{tab_sys_5}, these reaches
are lowered to $0.7\,{\rm TeV}$ and $1.0\,{\rm TeV}$ for $5\sigma$ and
$1.3\,{\rm TeV}$ and $1.8\,{\rm TeV}$ at $95\,\%$ C.L.  The experimental
reach is better for wino than Higgsino due to its large $SU(2)_L$
charge.  However, the wino DM could be as heavy as $2.9\,{\rm
TeV}$~\cite{ArkaniHamed:2006mb, Hisano:2006nn, Moroi:2013sla,
Beneke:2016ync} within the thermal freeze-out mechanism, which is beyond
our sensitivity.  On the other hand, for the Higgsino case, the mass of
$1.1\,{\rm TeV}$\cite{ArkaniHamed:2006mb, Cirelli:2007xd} favored by the
relic abundance is well within our reach at $95\,\%$ C.L. even if we
take account of the systematic errors.

We compare our results with those expected in the direct search at
future hadron colliders~\cite{Han:2018wus}.  Adopting the same beam
energy and the luminosity (i.e., $\sqrt{s}=100\, {\rm TeV}$ and
$\mathcal{L} = 30\,{\rm ab}^{-1}$), the Higgsino mass reaches at
$95\,\%$ C.L. are estimated as $m_{\tilde{H}} < 0.9 \hyphen 1.4\,{\rm TeV}$
by the mono-jet search (with $2\,\% \hyphen 1\,\%$ systematic errors) and
$m_{\tilde{H}} < 1.1 \hyphen 1.5\,{\rm TeV}$ by the disappearing track
search (with $500\,\% \hyphen 20\,\%$ uncertainty in background
estimation).\footnote{Note that the result of the disappearing track
  search described in~\cite{Han:2018wus} assumes a detector design
  similar to the ATLAS tracking system for the Run-2 $13\,{\rm TeV}$
  LHC with Insertable B-Layer~\cite{Capeans:1291633}.}  As we have
already mentioned, the disappearing track search is relevant only when
the mass difference of Higgsino is small enough.  In any case, the
indirect search based on the precision measurements gives comparable
or better sensitivity to Higgsino.  Also, we stress that the
five parameters fit used here is just an example and that we may be
able to discover Higgsino around $1\,{\rm TeV}$ by reducing the
systematic errors.  The situation is different for wino.  The mass
difference of the charged and the neutral components of wino tends to
be very small ($\Delta m \simeq 165\,{\rm MeV}$~\cite{Ibe:2012sx}).\footnote{
	In terms of the low energy effective theory,
	at least a dimension-seven operator is required
	to generate the mass difference for wino
	other than the electroweak correction~\cite{Gherghetta:1999sw}.
} Accordingly, the disappearing track search provides by far the most
efficient way for the wino search with the sensitivity up to $6\,{\rm
  TeV}$ \cite{Han:2018wus}.  Yet, of course, our analysis could
provide some independent information also for the wino search.

We also comment on the detection reach of the MDM scenario shown in
Fig.~\ref{subfig:57}.  The $5\sigma$ reaches are $3.9\,{\rm TeV}$ and
$2.2\,{\rm TeV}$ for 5-plet fermion and 7-plet scalar, while the
$95\,\%$ reaches are $6.2\,{\rm TeV}$ and $3.6\,{\rm TeV}$.  They are
lowered to $2.6\,{\rm TeV}$ and $1.2\,{\rm TeV}$ ($5\sigma$) and
$3.9\,{\rm TeV}$ and $2.1\,{\rm TeV}$ ($95\,\%$ C.L.) when the systematic
errors are included.  If we assume the vanilla thermal freeze-out
scenario, the mass should be $10\,{\rm TeV}$ for 5-plet fermion and
$25\,{\rm TeV}$ for 7-plet scalar~\cite{Cirelli:2007xd}.  Thus, our
method probes only a part of the allowed mass range for these
multiplets.

Next, in order to investigate how it is important to reduce the
systematic errors, we repeat the same procedure assuming that the
dominant sources of the systematic errors are well understood and
$\bm{\sigma}$ are all reduced by a factor of ten.  We obtain $5\sigma$
reach as $0.8\,{\rm TeV}$ and $1.1\,{\rm TeV}$ for Higgsino and wino,
both of which are improved by $100\,{\rm GeV}$.  Such an improvement is
expected if, in this example, we can reduce the errors from the
renormalization scale choice through the calculation of higher order
effects, and from those related to PDF by obtaining more profound
knowledge about PDF itself.  If we can reduce the systematic errors
further, the detection reach ultimately approaches to the optimistic
ones.

\begin{figure}[t]
 \centering
 \includegraphics[width=0.495\hsize]{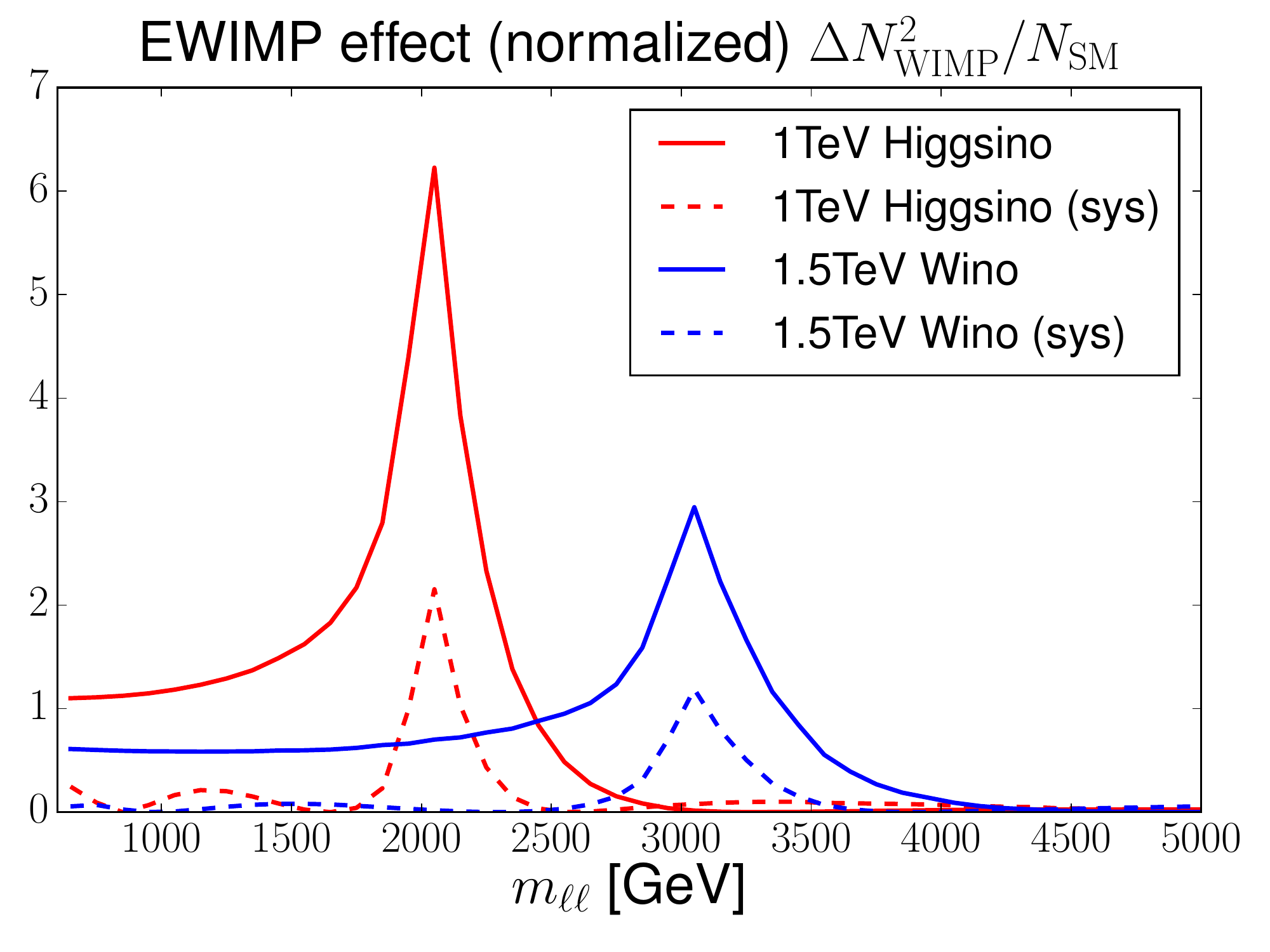}
 \caption{Plot of the contribution of each bin to the value of $q_0$.
 The color and line style conventions are the same as those in
 Fig.~\ref{subfig:wh}.}
 \label{fig_nlo2}
\end{figure}

Finally, in order to take a closer look at the significance of the
peak structure (Fig.~\ref{fig_sig_ratio}), we plot in
Fig.~\ref{fig_nlo2} the contribution of each bin to the value of
$q_0$.  The color and line style conventions are the same as those in
the previous figure.  We can see that the most contributions come from
the bins around the peak at $m_{\ell\ell} = 2m$.  This feature is
clearer for the fitting based approach, where all the smooth parts of
the correction are absorbed into the fit parameters, thus there is
almost no contribution to $q_0$ from the bins other than $m_{\ell\ell}
\sim 2m$.  Note also that, for the optimistic bound, there are more
contributions from the bins with lower $m_{\ell\ell}$ than those with
higher $m_{\ell\ell}$, though sometimes the cross section correction
is much larger in the latter bins.  This is just because of the
difference of number of events in each bin, that is
$\mathcal{O}(10^7)$ for $500\,{\rm GeV} < m_{\ell\ell} < 600\,{\rm
  GeV}$, while $\mathcal{O}(10^3)$ for $4900\,{\rm GeV} < m_{\ell\ell}
< 5000\,{\rm GeV}$ in our set up, for instance.

\subsection{Mass determination}
\label{sec_mass}

In this subsection, we briefly discuss the mass determination of the
EWIMP after its discover, which is also possible because of the
characteristic shape of the signal.  Here, we assume that the
underlying model is the SM $+$ EWIMP model with $\mu=1$, and calculate
the theoretical expectation of the number of events in each bin
$\tilde{x}_i$ as a function of the EWIMP mass $m$, i.e.  $\tilde{x}_i
= \tilde{x}_i(m)$. We further include the effects of the systematic
errors using $\tilde{x}_i(\bm{\theta}, m) \equiv \tilde{x}_i(m)
f_{\mathrm{sys},i}(\bm{\theta})$.  Since our purpose here is to
discuss the mass determination after the discovery, we assume that the
observed number of events is given by $\bm{x} =
\{\tilde{x}_i(m_\mathrm{true})\}$ in the following.  Then, the
accuracy of the determination of the EWIMP mass is estimated by the
following test statistic:
\begin{align}
  q_{m_\mathrm{fit}} 
 = -2 \ln \frac{L(\bm{x}; \doublehat{\bm{\theta}}, m_{\rm fit})}{L(\bm{x}; \hat{\bm{\theta}}, \hat{m})},\label{eq:qmfit}
\end{align}
where $\doublehat{\bm{\theta}}$ and $\{\hat{\bm{\theta}}, \hat{m}\}$ 
are the best fit values of $\bm{\theta}$ (and $m$) which maximize 
$L(\bm{x}; \bm{\theta}, m_\mathrm{fit})$ and $L(\bm{x}; \bm{\theta}, m)$, respectively.
The likelihood function is
\begin{align}
 L(\bm{x}; \bm{\theta}, m) &\equiv L_{\bm{\theta}} (\bm{x}; m)
 L'(\bm{\theta}, \bm{\sigma}),\\
 L_{\bm{\theta}} (\bm{x}; m) &\equiv
 \prod_{i} \exp\left[
 -\frac{(x_i - \tilde{x}_i(\bm{\theta}, m))^2}{2\tilde{x}_i(\bm{\theta}, m)}
 \right],\label{eq_l_m}
\end{align}
and $L'(\bm{\theta}, \bm{\sigma})$ given by Eq.~\eqref{eq_L_theta} with
$\bm{\sigma}$ being $\bm{0}$ for the optimistic case, and being the same
as those in Sec.~\ref{sec_statistical} for the case with the systematic
errors.

\begin{figure}[t]
 \centering
 \includegraphics[width=0.495\hsize]{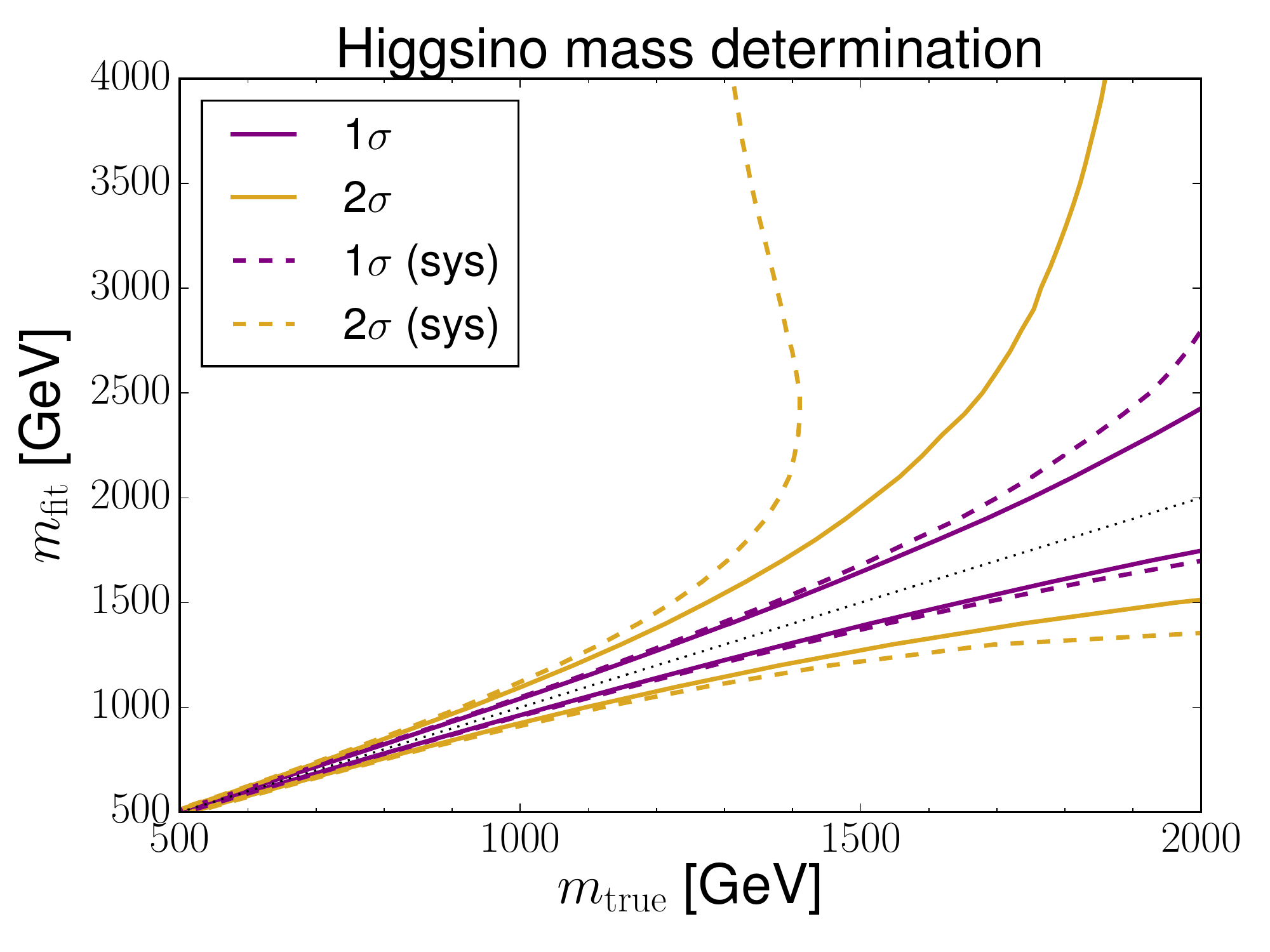}
 \caption{Contours of the test statistic for Higgsino in the $m_{\rm
 true}$ vs. $m_{\rm fit}$ plane.  The purple and golden lines
 correspond to $\sqrt{q_{m_\mathrm{fit}}} = 1$ and $2$, respectively.
 The dotted and solid lines are with and without the systematic
 errors.  For Higgsino within the $2\sigma$ discovery reach, these
 contours can be interpreted as $1\sigma$ and $2\sigma$ errors for the
 mass.}  \label{fig_mass_fit}
\end{figure}

In Fig.~\ref{fig_mass_fit}, we show contours of the test statistic for
Higgsino in the $m_{\rm true}$ vs. $m_{\rm fit}$ plane, from which we
can discuss the precision of the mass determination.  The purple and
golden lines correspond to $\sqrt{q_{m_\mathrm{fit}}} = 1$ and $2$,
respectively.  The dotted and solid lines are the results with and without
the systematic errors.  As shown in the figure, for light enough
Higgsino, the Higgsino mass can be determined with the precision of
$\pm (1\hyphen 100)\,{\rm GeV}$ at $1\sigma$ depending on the value of the
mass itself.  On the other hand, the upper bound on the Higgsino mass
becomes divergent when $m_\mathrm{true}$ is equal to the discovery
reach of the Higgsino at the corresponding C.L.  (see
Fig.~\ref{fig_chi2mass}), as expected from Eqs.~\eqref{q(theta)} and
\eqref{eq:qmfit}.\footnote{ For the case with systematic errors, the
  contours giving the upper bounds on the mass enter the regions above
  the discovery reach.  This is due to the fact that, with systematic
  errors, the value of $q_{m_\mathrm{fit}}$ is mostly determined by
  the bins around $m_\mathrm{true}$ and $m_\mathrm{fit}$.}  (Notice
that, as $m\rightarrow\infty$, $\delta_\sigma^{ab}
(m_{\ell\ell})\rightarrow 0$.)

Before closing this section, we comment here that the analysis
proposed here can be also used to discriminate the quantum number of
EWIMP.  For instance, we have checked that the signal due to the
existence of $1\,\mathrm{TeV}$ Higgsino cannot be well-fitted by wino,
5-plet fermion, nor 7-plet scalar irrespective of their masses.

\section{Conclusion}
\label{sec:conclusion}

We have studied the prospect of the indirect search of EWIMPs 
utilizing the Drell-Yan process
at future $100\,{\rm TeV}$ hadron colliders.  First, we performed an
optimistic analysis neglecting all the systematic errors, and obtained
the reach up to $2.0\,{\rm TeV}$ and $2.7\,{\rm TeV}$ at $95\,\%$
C.L. for Higgsino and wino, respectively.  Next, we have also taken
into account systematic errors by using a five parameters function fit
and seen that the limits are lowered to $1.3\,{\rm TeV}$ and $1.8\,{\rm
  TeV}$.  
These bounds for Higgsino are comparable to or better than
the other search strategies that make use of direct production of Higgsino at the colliders.
For wino, the
analysis proposed here can also provide some information independent
from the direct search.  We have also shown the $95\,\%$ C.L. reach
for 5-plet fermion and 7-plet scalar: $6.2\,{\rm TeV}$ and $3.6\,{\rm
  TeV}$ for the optimistic analysis and $3.9\,{\rm TeV}$ and
$2.1\,{\rm TeV}$ for the analysis with a fitting procedure.  The above
values can also be interpreted as the upper limit on the EWIMP masses
if we instead consider the exclusion prospect.

We have also applied our analysis to the determination of the EWIMP
mass after its discovery.  As an example, we have shown the result of
the Higgsino mass determination.  The Higgsino mass can be
reconstructed with an uncertainty of $\pm (1 \hyphen 100)\,{\rm GeV}$ at
the $1\sigma$ level, depending on the mass itself.

\section*{Acknowledgments}

This work was supported by JSPS KAKENHI Grant (Nos.\ 17J00813 [SC],
18J00540 [YE], 16H06490 [TM], and 18K03608 [TM].).

\bibliographystyle{elsarticle-num}
\bibliography{draft}

\end{document}